\documentclass[a4paper,11pt]{article}
\usepackage{amsmath}
\usepackage{amsfonts}
\usepackage{amssymb}
\newtheorem{theorem}{Theorem}

\newtheorem{definition}[theorem]{Definition}
\newtheorem{lemma}[theorem]{Lemma}
\newtheorem{proposition}[theorem]{Proposition}
\newenvironment{proof}[1][Proof]{\textbf{#1.} }{\ \rule{0.5em}{0.5em}}

\numberwithin{equation}{section}

\begin{document}

\title{\textbf{Generalized WDVV equations for }$B_{r}$ \textbf{and }$C_{r}$\textbf{%
\ pure N=2 Super-Yang-Mills theory}}
\date{}
\author{L.K. Hoevenaars, R. Martini}
\maketitle

\begin{abstract}
\noindent
A proof that the prepotential for pure N=2 Super-Yang-Mills theory
associated with Lie algebras $B_{r}$ and $C_{r}$ satisfies the generalized
WDVV (Witten-Dijkgraaf-Verlinde-Verlinde) system was given by
Marshakov, Mironov and Morozov. Among other things, they use an
associative algebra of holomorphic differentials.
Later Ito and Yang used a different approach to try to accomplish the
same result, but they encountered objects of which it is unclear whether they
form structure constants of an associative algebra.
We show by explicit calculation that these objects are none other
than the structure constants of the algebra of holomorphic differentials.
\end{abstract}

\section{Introduction}

In 1994, Seiberg and Witten \cite{SEIB-WITT1:1994}\ solved the low energy
behaviour of pure N=2 Super-Yang-Mills theory by giving the solution of the
prepotential $\mathcal{F}$. The essential ingredients in their construction
are a family of Riemann surfaces $\Sigma$, a meromorphic differential
$\lambda_{SW}$\ on it and the definition of the prepotential in terms of
period integrals of $\lambda_{SW}$%
\begin{equation}
a_{i}=\int_{A_{i}}\lambda_{SW}\text{ \ \ \ \ \ \ \ }\frac{\partial\mathcal{F}%
}{\partial a_{i}}=\int_{B_{i}}\lambda_{SW}%
\end{equation}
where $A_{i}$ and $B_{i}$ belong to a subset of the canonical cycles on the
surface $\Sigma$ and the $a_{i}$ are a subset of the moduli parameters of the
family of surfaces. These formulas define the prepotential $\mathcal{F}\left(
a_{1},...,a_{r}\right)  $ implicitly; here $r$ denotes the rank of the gauge
group under consideration.

\bigskip
\noindent 
A link between the prepotential and the Witten-Dijkgraaf-Verlinde-Verlinde
equations \cite{WITT:1991},\cite{DIJK-VERL-VERL:1991} was first suggested in
\cite{BONE-MATO:1996}. Since then an extensive literature on the subject was
formed. It was found that the perturbative piece of the prepotential
$\mathcal{F}(a_{1},...,a_{r})$ for pure N=2 SYM theory satisfies the
generalized WDVV equations \cite{MARS-MIRO-MORO:1996},\cite{MART-GRAG:1999}%
,\cite{VESE:1999}
\begin{equation}
\mathcal{F}_{i}\mathcal{F}_{k}^{-1}\mathcal{F}_{j}=\mathcal{F}_{j}%
\mathcal{F}_{k}^{-1}\mathcal{F}_{i}\text{ \ \ \ \ }\forall i,j,k=1,...,r
\end{equation}
where the $\mathcal{F}_{i}$ are matrices given by $\left(  \mathcal{F}%
_{i}\right)  _{jk}=\frac{\partial^{3}\mathcal{F}}{\partial a_{i}\partial
a_{j}\partial a_{k}}$.

\noindent
Moreover, it was shown that the full prepotential for simple Lie algebras of
type A,B,C,D \cite{MARS-MIRO-MORO:2000} and type E
\cite{ITO-YANG:1998} and F \cite{HOEV-KERS-MART:2000} satisfies this
generalized WDVV system\footnote{The generalized WDVV equations for
  $G_2$ are trivial}. The approach used by Ito and Yang in \cite{ITO-YANG:1998}
differs from the other two, due to the type of associative algebra
that is being used: they use the Landau-Ginzburg chiral ring while the 
others use an algebra of holomorphic differentials. For the A,D,E
cases this difference in approach is negligible since the two
different types of algebras are isomorphic. For the Lie
algebras of B,C type this is not the case and this leads to some
problems. The present article deals with these problems and shows that 
the proper algebra to use is the one suggested in \cite{MARS-MIRO-MORO:2000}. 
A survey of these matters, as well as the results of the
present paper can be found in the internal publication \cite{HOEV-MART:2000}.

\bigskip
\noindent
This paper is outlined as follows: in the first section we will review 
Ito and Yang's method for the A,D,E Lie algebras. In the second
section their approach to B,C Lie algebras is discussed. Finally in
section three we show that Ito and Yang's construction naturally
leads to the algebra of holomorphic differentials used in
\cite{MARS-MIRO-MORO:2000}.

\section{A review of the simply laced case}

In this section, we will describe the proof in \cite{ITO-YANG:1998} 
that the prepotential of 4-dimensional pure $N=2$ SYM theory
with Lie algebra of simply laced (ADE) type satisfies the generalized WDVV
system. The Seiberg-Witten data \cite{SEIB-WITT1:1994}, \cite
{GORS-KRIC-MARS-MIRO-MORO:1995}, \cite{MART-WARN:1996} consists of:

\bigskip

\begin{itemize}
\item  a family of Riemann surfaces $\Sigma$ of genus $g$ given by 
\begin{equation}
z+\dfrac{\mu }{z}=W(x,u_{1},\ldots ,u_{r})  \label{surface_ADE}
\end{equation}
where $W$ is a particular function which is polynomial for Lie algebras $%
A_{r}$, rational for $D_{r}$ and contains some radicals for the exceptional
algebras. Furthermore $\mu $ is a scaling parameter which plays a role in
the physical theory, $r$ is the rank of the Lie algebra and the $u_{i}$ its
Casimirs, which play the role of moduli parameters for the family of
surfaces.

\item  a meromorphic differential $\lambda _{SW}$ on the surfaces,
called the Seiberg-Witten differential. It is defined by 
\begin{equation}
\lambda _{SW}=x\dfrac{dz}{z}
\end{equation}

and has the property that $\dfrac{\partial \lambda _{SW}}{\partial u_{i}}$
is a holomorphic differential modulo exact forms.

\item  a particular subset $\{A_{i},B_{i}\}$ containing $2r$ of a set of $2g$
canonical homology cycles.
\end{itemize}

\bigskip

\bigskip

\noindent From this data one can implicitly construct the prepotential by
first defining 
\begin{equation}
a_{i}:=\oint_{A_{i}}\lambda _{SW}\qquad \mathcal{F}_{j}:=\oint_{B_{j}}%
\lambda _{SW}
\end{equation}
and noting that the period matrix $\dfrac{\partial \mathcal{F}_{j}}{\partial
a_{i}}$ is symmetric. This implies that $\mathcal{F}_{j}$ can be thought of
as a gradient, which leads to the following

\begin{definition}
The prepotential is a function $\mathcal{F}(a_{1},\ldots ,a_{r})$ such that 
\begin{equation}
\mathcal{F}_{j}=\dfrac{\partial \mathcal{F}}{\partial a_{j}}
\end{equation}
\end{definition}

\noindent We also define a system of third order nonlinear partial
differential equations:

\begin{definition}
Let $f:\mathbb{C}^{r}\rightarrow \mathbb{C}$, then the generalized WDVV
system \cite{BONE-MATO:1996}, \cite{MARS-MIRO-MORO:1996} for $f$ is 
\begin{equation}
f_{i}K^{-1}f_{j}=f_{j}K^{-1}f_{i}\quad \forall i,j\in \left\{
1,...,r\right\} 
\end{equation}
where the $f_{i}$ are matrices with entries 
\begin{equation}
(f_{i})_{jk}=\dfrac{\partial ^{3}f(a_{1},...,a_{r})}{\partial a_{i}\partial
a_{j}\partial a_{k}}
\end{equation}
and $K=\sum_{l=1}^{r}\alpha _{l}f_{l}$ is an invertible linear combination
of them.
\end{definition}

\bigskip

\noindent The main result of this section is

\begin{proposition}
The prepotential satisfies the generalized WDVV system
\end{proposition}

\begin{proof}
We will prove that the prepotential satisfies this system by showing
existence of the following:

\begin{enumerate}
\item  an `invertible metric' $K$

\item  an associative algebra with structure constants $C_{ij}^{k}$

\item  a relation between the third order derivatives of $\mathcal{F}$ and
the structure constants: 
\begin{equation}
(\mathcal{F}_{i})_{jk}=C_{ij}^{l}K_{kl}  \label{relation}
\end{equation}
\end{enumerate}

\begin{lemma}
\textit{If conditions }$1-3$\textit{\ are met, then }$F$\textit{\ satisfies
the generalized WDVV system.}
\end{lemma}

\begin{proof}
\textit{Associativity of the algebra can be expressed through an identity on
the structure constants} 
\begin{equation}
C_{i}C_{j}=C_{j}C_{i}
\end{equation}
\textit{which due to 3 also reads } 
\begin{equation}
\mathcal{F}_{i}K^{-1}\mathcal{F}_{j}K^{-1}=\mathcal{F}_{j}K^{-1}\mathcal{F}%
_{i}K^{-1}
\end{equation}
\textit{and multiplying by }$K$\textit{\ from the right gives the
desired result.}
\end{proof}

\ \linebreak \noindent The rest of the proof deals with a discussion of the
conditions 1-3. It is well-known \cite{KLEM-LERC-YANK-THEI:1995} that the
right hand side of (\ref{surface_ADE}) equals the Landau-Ginzburg
superpotential associated with the corresponding Lie algebra. Using this
connection, we can define the primary fields $\phi _{i}(u):=-\dfrac{\partial
W}{\partial u_{i}}$ and

\begin{definition}
The chiral ring is an associative algebra defined by 
\begin{equation}
\phi _{i}(u)\phi _{j}(u)=C_{ij}^{k}(u)\phi _{k}(u)\qquad \mbox{mod}\left( 
\dfrac{\partial W}{\partial x}\right) 
\end{equation}
\end{definition}

\noindent Instead of using the $u_{i}$ as coordinates on the part of
the moduli space we're interested in, we
want to use the $a_{i}$. For the chiral ring this implies that in the new
coordinates 
\begin{eqnarray}
(-\dfrac{\partial W}{\partial a_{i}})(-\dfrac{\partial W}{\partial a_{j}}) & 
= & \dfrac{\partial u_{x}}{\partial a_{i}}\dfrac{\partial u_{y}}{\partial
a_{j}}C_{xy}^{z}(u)\dfrac{\partial a_{k}}{\partial u_{z}}(-\dfrac{\partial W%
}{\partial a_{k}})\qquad \mbox{mod}(\dfrac{\partial W}{\partial x})
\qquad \Longrightarrow \nonumber \\ \nonumber \\
\phi _{i}(a)\phi _{j}(a) & = & C_{ij}^{k}(a)\phi _{k}(a)\qquad \mbox{mod}(%
\dfrac{\partial W}{\partial x})
\end{eqnarray}
which again is an associative algebra, but with different structure
constants $C_{ij}^{k}(a)\neq C_{ij}^{k}(u)$. This is the algebra we will use
in the rest of the proof.\newline

\noindent For the relation (\ref{relation}) we turn to another aspect of
Landau-Ginzburg theory: the Picard-Fuchs equations (see e.g \cite
{LERC-SMIT-WARN:1992} and references therein). These form a coupled set of
first order partial differential equations which express how the integrals
of holomorphic differentials over homology cycles of a Riemann surface in a
family depend on the moduli.

\begin{definition}
Flat coordinates of the Landau-Ginzburg theory are a set of coordinates $%
\{t_{i}\}$ on moduli space such that 
\begin{equation}
\dfrac{\partial ^{2}W}{\partial t_{i}\partial t_{j}}=\dfrac{\partial Q_{ij}}{%
\partial x}
\end{equation}
where $Q_{ij}$ is given by 
\begin{equation}
\phi _{i}(t)\phi _{j}(t)=C_{ij}^{k}(t)\phi _{k}(t)+Q_{ij}\dfrac{\partial W}{%
\partial x}
\end{equation}
\end{definition}

\noindent The fact that such coordinates indeed exist will not be discussed
here. In terms of these coordinates the following set of Picard-Fuchs
equations hold \cite{ITO-YANG:1997} 
\begin{equation}
\dfrac{\partial }{\partial t_{i}}\left[ \oint_{\Gamma }\dfrac{\partial
\lambda sw}{\partial t_{j}}\right] =C_{ij}^{k}(t)\dfrac{\partial }{\partial
t_{k}}\left[ \oint_{\Gamma }\dfrac{\partial \lambda sw}{\partial t_{r}}%
\right]   \label{PF_t}
\end{equation}
for any cycle $\Gamma \in \{A_{i},B_{i}\}$. These equations were derived by
making use of the chiral ring, expressed in the flat coordinates, and
therefore the structure constants $C_{ij}^{k}(t)$ appear. Making a change of
coordinates to the $a_{i}$ and using the fact that the $a_{i}$ satisfy (\ref
{PF_t}) one finds 
\begin{equation}
\dfrac{\partial }{\partial a_{i}}\left[ \oint_{\Gamma }\dfrac{\partial
\lambda sw}{\partial a_{j}}\right] -C_{ij}^{l}(a)\dfrac{\partial }{\partial
a_{l}}\left[ \oint_{\Gamma }\dfrac{\partial \lambda sw}{\partial a_{m}}%
\right] \dfrac{\partial a_{m}}{\partial t_{r}}
\end{equation}

\noindent Taking $\Gamma =B_{k}$ we get 
\begin{equation}
\mathcal{F}_{i}{}_{jk}=C_{ij}^{l}(a)K_{kl}
\end{equation}
which is the intended relation (\ref{relation}). The only thing that is left
to do, is to prove that $K_{kl}=\dfrac{\partial a_{m}}{\partial t_{r}}%
\mathcal{F}_{mkl}$ is invertible. This will not be discussed in the present
paper.
\end{proof}

\ \newline
\noindent In conclusion, the most important ingredients in the proof are the
chiral ring and the Picard-Fuchs equations. In the following sections we
will show that in the case of $B_{r},C_{r}$ Lie algebras, the Picard-Fuchs
equations can still play an important role, but the chiral ring should be
replaced by the algebra of holomorphic differentials considered by
the authors of \cite{MARS-MIRO-MORO:2000}. These algebras
are isomorphic to the chiral rings in the ADE cases, but not for Lie
algebras $B_{r},C_{r}$.

\section{Ito \& Yang's approach to $B_{r}$ and $C_{r}$}

In this section, we discuss the attempt made in \cite{ITO-YANG:1998} to
generalize the contents of the previous section to the Lie algebras $%
B_{r},C_{r}$. We will discuss only $B_{r}$ since the situation for $C_{r}$
is completely analogous. The Riemann surfaces are given by 
\begin{equation}
z+\dfrac{\mu }{z}=\dfrac{W_{BC}(x,u_{i})}{x}  \label{surface_B}
\end{equation}
where $W_{BC}$ is the Landau-Ginzburg superpotential associated with the
theory of type $BC$. From the superpotential we again construct the chiral
ring in flat coordinates where 
\begin{equation}
\phi _{i}(t):=-\dfrac{\partial W_{BC}}{\partial t_{i}}\qquad \qquad \phi
_{i}(t)\phi _{j}(t)=C_{ij}^{k}(t)\phi _{k}(t)+Q_{ij}\left[ \dfrac{\partial
W_{BC}}{\partial x}\right] 
\label{chiralB}
\end{equation}
However, the fact that the right-hand side of (\ref{surface_B}) does not
equal the superpotential is reflected by the Picard-Fuchs equations, which
no longer relate the third order derivatives of $\mathcal{F}$ with the
structure constants $C_{ij}^{k}(a)$. Instead, they read 
\begin{equation}
\mathcal{F}_{ijk}=\tilde{C}_{ij}^{l}(a)K_{kl}
\end{equation}
where $K_{kl}=\dfrac{\partial a_{m}}{\partial t_{r}}\mathcal{F}_{mkl}$ and 
\begin{equation}
\tilde{C}_{ij}^{k}(t)=C_{ij}^{k}(t)+D_{ij}^{l}\sum_{n=1}^{r}\dfrac{2nt_{n}}{%
2r-1}\tilde{C}_{nl}^{k}(t).  \label{Ctilde}
\end{equation}
The $D_{ij}^{l}$ are defined by 
\begin{equation}
Q_{ij}=xD_{ij}^{l}\phi _{l}  \label{D}
\end{equation}
and we switched from $\tilde{C}_{ij}^{k}(a)$ to $\tilde{C}_{ij}^{k}(t)$ in order to compare these with
the structure constants $C_{ij}^{k}(t)$. At this point, it is unknown%
\footnote{%
Except for rank 3 and 4, for which explicit calculations of $\tilde{C}%
_{ij}^{k}(t)$ were made in \cite{ITO-YANG:1998}} whether the $\tilde{C}%
_{ij}^{k}(t)$ (and therefore the $\tilde{C}_{ij}^{k}(a)$) are structure
constants of an associative algebra. This issue will be resolved in the next
section.

\section{The identification of the structure constants}

The method of proof that is being used in \cite{MARS-MIRO-MORO:2000}
for the $B_{r},C_{r}$ case also involves an associative algebra. However,
theirs is an algebra of holomorphic differentials which is isomorphic to 
\begin{equation}
\phi _{i}(t)\phi _{j}(t)=\gamma _{ij}^{k}(t)\phi _{k}(t)\qquad \mbox{mod}(x%
\dfrac{\partial W_{BC}}{\partial x}-W_{BC}).
\end{equation}
In the rest of this section we will show that

\begin{theorem}
$\qquad \qquad \qquad \qquad \qquad $\fbox{$\tilde{C}_{ij}^{k}(t)=\gamma
_{ij}^{k}(t)$}
\end{theorem}

\begin{proof}
Starting from the multiplication structure in (\ref{chiralB}) we find (see (\ref{D})) 
\begin{equation}
\phi _{i}(t)\phi _{j}(t)=\sum_{k=1}^{r}C_{ij}^{k}(t)\phi
_{k}(t)+\sum_{k=1}^{r}D_{ij}^{k}\phi _{k}(t)x\partial _{x}W_{BC}
\label{algebra with D}
\end{equation}
we will rewrite it in such a way that it becomes of the form 
\begin{equation}
\phi _{i}(t)\phi _{j}(t)=\sum_{k=1}^{r}\widetilde{C}_{ij}^{k}(t)\phi
_{k}(t)+P_{ij}\left[ x\partial _{x}W_{BC}-W_{BC}\right] 
\end{equation}
As a first step, we use (\ref{Ctilde}): 
\begin{eqnarray}
\phi _{i}\phi _{j} &=&\left[ C_{i}\cdot \overrightarrow{\phi }+D_{i}\cdot 
\overrightarrow{\phi }x\partial _{x}W_{BC}\right] _{j}  \notag \\
&=&\left[ \left( \widetilde{C}_{i}-D_{i}\cdot \sum_{n=1}^{r}\frac{2nt^{n}}{%
2r-1}\widetilde{C}_{n}\right) \cdot \overrightarrow{\phi }+D_{i}\cdot 
\overrightarrow{\phi }x\partial _{x}W_{BC}\right] _{j}  \notag \\
&=&\left[ \widetilde{C}_{i}\cdot \overrightarrow{\phi }-D_{i}\cdot
\sum_{n=1}^{r}\frac{2nt^{n}}{2r-1}\widetilde{C}_{n}\cdot \overrightarrow{%
\phi }+D_{i}\cdot \overrightarrow{\phi }x\partial _{x}W_{BC}\right]
_{j}
\label{second term}
\end{eqnarray}
The notation $\overrightarrow{\phi }$ stands for the vector with components $%
\phi _{k}$ and we used a matrix notation for the structure constants.

\noindent The proof becomes somewhat technical, so let us first
give a general outline of it. The strategy will be to get rid of the second
term of (\ref{second term}) by cancelling it with part of the third term,
since we want an algebra in which the first term gives the structure
constants. For this cancelling we'll use equation $\left( \ref{Ctilde}%
\right) $ in combination with the following relation which expresses the
fact that $W_{BC}$ is a graded function
\begin{equation}
x\frac{\partial W_{BC}}{\partial x}+\sum_{n=1}^{r}2nt_{n}\frac{\partial
W_{BC}}{\partial t_{n}}=2rW_{BC}  \label{scaling}
\end{equation}
Cancelling is possible at the expense of introducing yet another term which
then has to be canceled etcetera. This recursive process does come to an end
however, and by performing it we automatically calculate modulo $x\partial
_{x}W_{BC}-W_{BC}$ instead of $x\partial _{x}W_{BC}$.
  
\qquad  \newline
\noindent
We rewrite $\left( \ref{second term}\right) $ by splitting up the third term
and rewriting one part of it using $\left( \ref{scaling}\right) $: 
\begin{eqnarray}
\left[ D_{i}\cdot \overrightarrow{\phi }x\partial _{x}W_{BC}\right] _{j} &=&%
\left[ -\frac{1}{2r-1}D_{i}\cdot \overrightarrow{\phi }x\partial
_{x}W_{BC}+\left( 1+\frac{1}{2r-1}\right) D_{i}\cdot \overrightarrow{\phi }%
x\partial _{x}W_{BC}\right] _{j}  \notag \\
&=&\left[ -\frac{D_{i}}{2r-1}\cdot \overrightarrow{\phi }\left(
2rW_{BC}-\sum_{n=1}^{r}2nt_{n}\phi _{n}\right) +\frac{2rD_{i}}{2r-1}\cdot 
\overrightarrow{\phi }x\partial _{x}W_{BC}\right] _{j}
\end{eqnarray}
Now we use (\ref{algebra with D}) to work out the product $\phi _{k}\phi _{n}
$ and the result is: 
\begin{eqnarray}
\phi _{i}\phi _{j}&=&\left[ \widetilde{C}_{i}\cdot \overrightarrow{\phi }-%
\frac{D_{i}}{2r-1}\cdot \sum_{n=1}^{r}2nt_{n}\left( \widetilde{C}_{n}\cdot 
\overrightarrow{\phi }-C_{n}\cdot \overrightarrow{\phi }\right) -\frac{D_{i}%
}{2r-1}\cdot \sum_{n=1}^{r}2nt_{n}D_{n}\cdot \overrightarrow{\phi }x\partial
_{x}W_{BC}\right] _{j} \notag \\
&+& \frac{2rD_{i}}{2r-1}\cdot \left[ x\partial _{x}W_{BC}-W_{BC}\right] _{j} 
\end{eqnarray}
We now use (\ref{Ctilde}) again to rewrite the second term in the first
line: 
\begin{eqnarray}
\phi _{i}\phi _{j} &=&\left[ \widetilde{C}_{i}\cdot \overrightarrow{\phi }+%
\frac{D_{i}}{2r-1}\cdot \sum_{n=1}^{r}2nt_{n}\left( -D_{n}\cdot
\sum_{m=1}^{r}\frac{2mt^{m}}{2r-1}\widetilde{C}_{m}\cdot \overrightarrow{%
\phi }+D_{n}\cdot \overrightarrow{\phi }x\partial _{x}W_{BC}\right) \right]
_{j}  \notag  \\
&&+\frac{2rD_{i}}{2r-1}\left[ x\partial _{x}W_{BC}-W_{BC}\right] _{j}
\end{eqnarray}
Note that by cancelling the one term, we automatically calculate modulo $%
x\partial _{x}W_{BC}-W_{BC}$. The expression between brackets in the first line seems to
spoil our achievement but it doesn't: until now we rewrote 
\begin{equation}
\left[ -D_{i}\cdot \sum_{n=1}^{r}\frac{2nt_{n}}{2r-1}\widetilde{C}_{n}\cdot 
\overrightarrow{\phi }+D_{i}\cdot \overrightarrow{\phi }x\partial _{x}W_{BC}%
\right] _{j}
\end{equation}
and we can now rewrite using the same procedure
\begin{equation}
\left[ -D_{n}\cdot \sum_{m=1}^{r}\frac{2mt^{m}}{2r-1}\widetilde{C}_{m}\cdot 
\overrightarrow{\phi }+D_{n}\cdot \overrightarrow{\phi }x\partial _{x}W_{BC}%
\right] _{j}
\end{equation}
This is a recursive process. If it stops at some point, then we get a
multiplication structure 
\begin{equation}
\phi _{i}\phi _{j}=\sum_{k=1}^{r}\widetilde{C}_{ij}^{k}\phi
_{k}+P_{ij}\left( x\partial _{x}W_{BC}-W_{BC}\right) 
\label{we willen deze algebra}
\end{equation}
for some polynomial $P_{ij}$ and the theorem is proven. To see that the
process indeed stops, we refer to the lemma below.
\end{proof}

\noindent
We note that the recursive process adds a matrix $D_{i}
$ each time, and these matrices have a pleasant property:
\begin{lemma}
The matrices $D_{i}$ are nilpotent.
\end{lemma}

\begin{proof}
One finds \cite{ITO-YANG:1998} that the degree of $Q_{ij}$ is 
\begin{equation}
\deg \left( Q_{ij}\right) =2r+1-2\left( i+j\right) 
\end{equation}
Dividing this by $x$, we get an object of degree $2r-2\left( i+j\right) $.
The $D_{ij}^{k}$ are defined through $\left( \ref{D}\right) \ $and if we can
show that for $j\geq k$ we can't divide $\frac{Q_{ij}}{x}$ by $\phi _{k}$,
we have shown that $D_{i}$ is nilpotent since it is strictly upper
triangular. Since 
\begin{equation}
\deg \left( \phi _{k}\right) =2r-2k
\end{equation}
we find that indeed for $j\geq k$ the degree of $\phi _{k}$ is bigger than
the degree of $\frac{Q_{ij}}{x}$ and since they are both polynomials we
can't divide the two. This finishes the proof of the lemma.
\end{proof}

\section{Conclusions and outlook}

In this letter we have shown that the unknown quantities $\tilde{C}_{ij}^{k}$
of \cite{ITO-YANG:1998} are none other than the structure constants of the
algebra of holomorphic differentials introduced in \cite{MARS-MIRO-MORO:2000}%
. Therefore this is the algebra that should be used, and not the
Landau-Ginzburg chiral ring. However, the connection with Landau-Ginzburg
can still be very useful since the Picard-Fuchs equations may serve as an
alternative to the residue formulas considered in \cite{MARS-MIRO-MORO:2000}.

\bigskip

\bibliographystyle{h-physrev}
\bibliography{biblio}

\end{document}